\documentclass[%
 reprint,
 amsmath,amssymb,
 aps,
 prd
]{revtex4-2}

\usepackage{graphicx}
\usepackage{xcolor}
\usepackage{placeins}
\usepackage{dcolumn}
\usepackage{bm}
\usepackage[hidelinks]{hyperref}

\begin{document}

\title{Tidal Heating of Stellar Clusters in Fuzzy Dark Matter Halos}

\author{Yiheng Liu}
\email{liuyiheng26@mails.tsinghua.edu.cn}
\author{Xinyu Li}
\email{xinyuli@tsinghua.edu.cn}
\affiliation{
 Department of Astronomy, Tsinghua University, 30 Shuangqing Rd, Beijing, 100084, China
}

\begin{abstract}
Ultra-faint dwarf galaxies serve as powerful testing grounds for wave dark matter models through dynamical stellar heating. Previous simulation-based work derived a lower bound on the fuzzy dark matter particle mass using a diffusion approximation valid only when the de Broglie wavelength is much smaller than the galaxy’s half-light radius. We simulate the dynamical evolution of stellar clusters in FDM halos across a wide mass range and find that for sufficiently low masses, where the de Broglie wavelength is much larger than the cluster size, tidal heating is the main mechanism. We also find that a reduced soliton mass and tidally stripped halo can suppress the heating. We demonstrate that in order to constrain FDM mass from cluster heating, the structure and environment of the FDM halo must be carefully considered.
\end{abstract}

\maketitle


\section{Introduction}

Fuzzy dark matter (FDM), also known as ultra-light dark matter or wave dark matter, has emerged as a well-motivated alternative model that naturally addresses these small-scale issues of the cold dark matter paradigm. In the FDM model, the dark matter is composed of an ultra-light bosonic field—often identified with the QCD axion or an axion-like particle arising from string compactifications—with a particle mass $10^{-21}-10^{-23}$~eV \cite{Hu2000,Hui2021}. The extraordinarily small particle mass gives rise to a macroscopic de Broglie wavelength $\lambda_\text{dB}=h/(m_a v)$ on galactic scales, which leads to the formation of extended solitonic cores in halo centers \cite{Schive2014,Schive2014b,Mocz2017,Li2019}.
The halos formed by FDM exhibit novel dynamical phenomena including ``granules'' of order-unity density fluctuations throughout the halo \cite{Mocz2017,Li2019}, formation of vortex lines \cite{Hui2021b} and random walk of the soliton cores \cite{Schive2020,Li2021}.
All these features are consequences of the wave interference.

The de Broglie length scale density fluctuations and random walk of the central soliton both lead to small-scale potential fluctuations which can dynamically heat the stars in clusters.
The heating rate is usually derived assuming stars randomly scattering off the de Broglie size quasi-particle. 
The growth of the velocity dispersion is derived using a diffusion formalism \cite{Hui2017,dalal2022}.
\begin{align}
\Delta\sigma_\star^2\simeq9\left(\frac{\sigma_\star}{\sigma_\text{dm}}\right)^4\left(\frac{\hbar}{m_a}\right)^3\frac{t}{r_{1/2}^4}.
\label{eq:heating}
\end{align}

Ultra-faint dwarf (UFD) galaxies such as Segue 1 and Segue 2 are ideal testbeds: they are dark-matter-dominated, have small 3D half-light radii $r_{1/2}\sim50\,\mathrm{pc}$ and low stellar velocity dispersions $\sigma_\star\approx2-5\,\mathrm{km/s}$.
\cite{dalal2022} studied heating of Segue 1 and Segue 2 in FDM halos.
Plugging in typical UFD parameters ($t=10\,\mathrm{Gyr}$, $r_{1/2}=50\,\mathrm{pc}$, $\sigma_\star=3\,\mathrm{km/s}$, $\sigma_\text{dm}=6\,\mathrm{km/s}$) into Equation~\ref{eq:heating} gives a characteristic mass scale $m_a\sim10^{-19}\,\mathrm{eV}$ where the accumulated heating becomes comparable to the observed velocity dispersion.
They derived a lower bound $m_a\gtrsim3\times10^{-19}\,\mathrm{eV}$ at $99\%$ confidence, effectively excluding the canonical FDM mass range.
Subsequent work has challenged this bound on several grounds: tidal stripping of the dark matter halo may remove the granular structure \cite{Dutta2023, Yang2025}, and the quasi-particle approximation itself may break down \cite{Widrow2024}. 
In response, \cite{May2025} updated their constraints using new numerical simulations, though still within the diffusive regime.

A crucial but neglected point is that Equation~\ref{eq:heating} is only valid when the stellar cluster size is much larger than the de Broglie wavelength.
Under this condition, different stars scatter off independent quasi-particles, and heating is diffusive. We call this ``diffusive heating''. When the particle mass $m_a$ is sufficiently small, the de Broglie wavelength becomes comparable to or larger than the galaxy size, $\lambda_\text{dB}\gtrsim r_{1/2}$. In this long-wavelength regime, the granular structure cannot be fully realized within the stellar system; the wave field inside the galaxy behaves more like a coherent, oscillating background rather than a collection of independent scatterers. Consequently, the heating mechanism should transition from diffusive to a fundamentally different regime known as tidal heating \cite{Gnedin1999}.
This transition has not been systematically explored in previous work.

In this paper, we use numerical simulations to systematically investigate stellar cluster heating in the long-wavelength regime, where $\lambda_{\mathrm{dB}}$ is much larger than the stellar cluster size. We show that tidal heating becomes the dominant mechanism in this regime, leading to qualitatively different behavior: exponential growth of cluster size and velocity dispersion, rather than diffusive $\sqrt{t}$ growth. This transition fundamentally alters the predicted heating efficiency at low $m_a$.

\section{Tidal Heating}
When the de Broglie wavelength $\lambda_{\mathrm{dB}}$ is much larger than the size of a stellar cluster, the cluster experiences the FDM halo’s potential as a smooth, coherent field across its extent. However, the potential still fluctuates in time due to the underlying wave interference. These time-varying tidal fields can transfer energy to the internal motions of the cluster — a process known as tidal heating \cite{Gnedin1999}.

Consider a stellar cluster moving in the satellite FDM halo.
Let $\mathbf{R}$ denote the position of its center of mass and $\mathbf{r}$ the displacement of a star from the center of mass.  
Neglecting self-gravity (valid for a low-mass cluster), the equation of motion for a star is \cite{Gnedin1999}
\begin{equation}
    \ddot{\mathbf{r}} = -\frac{\mathrm{d}^2\Phi}{\mathrm{d}\mathbf{R}^2}\cdot\mathbf{r}.
\end{equation}
where $\Phi$ is the gravitational potential. 
We consider the situation that the satellite halo dominates the gravitational potential.
This equation describes how the tidal field, the second derivative of the potential, stretches the cluster.

In the satellite FDM halo, the potential fluctuates on a characteristic scale $\lambda_{\mathrm{dB}}/2$ with amplitude $\Phi_0$.
Here we consider a sinusoidal wave, so the over-dense/under-dense region each covers half the wavelength.
The magnitude of the tidal tensor can be approximated as
\begin{equation}
    \ddot{r} \sim \left|\frac{4\Phi_0}{\lambda_{\mathrm{dB}}^2}\right|r.
\end{equation}
This is an exponential growth equation: $r \propto \exp(k_R t)$, with growth rate $k_R\sim 2\sqrt{\Phi_0}/\lambda_{\mathrm{dB}}$.
At the satellite halo center, a soliton is formed with mass $M_c$ and radius $r_c$. 
The de Broglie wavelength can be estimated as $\lambda_{\mathrm{dB}}\sim h/m\sqrt{\Phi_0}$
where the potential is given by $\Phi_0\sim G M_c/r_c$.
Therefore the exponential growth rate
\begin{equation}\label{growthrate}
    k_R\sim \frac{GM_c}{2h r_c}m_a
\end{equation}
is proportional to $m_a$.
The kinetic energy of the cluster grows as $r^2$ \cite{Gnedin1999}, so the velocity dispersion $\sigma_v$ will also grow exponentially $\sigma\sim \exp{(k_\sigma t)}$ with $k_\sigma\sim k_R$ if the cluster remains spherical.

Several important points deserve emphasis here. 
First, tidal heating arises from the potential fluctuation produced by the granular density field, but with amplitude reduced, as stars in the cluster only ``see'' the same de Broglie size blob.  
Second, the exponential growth rate is proportional to $m_a$, meaning lower-mass FDM particles produce slower heating. 
Third, tidal heating is only valid while the cluster size $r \ll \lambda_{\mathrm{dB}}$. 
Once the cluster expands to become comparable to $\lambda_{\mathrm{dB}}$, the approximation that the tidal field is uniform across the cluster breaks down. 
At that point, the cluster can no longer be treated as a point-like object relative to the granularity scale, and the heating mechanism transitions to the diffusive regime. 

Thus, for a given $m_a$, we expect the following evolutionary sequence: an initial exponential expansion phase (tidal heating) when $r \ll \lambda_{\mathrm{dB}}$, then transition to a subsequent diffusive heating phase ($\sigma \propto \sqrt{t}$) for $r \gg \lambda_{\mathrm{dB}}$ until it is virialized with the satellite halo. The transition time is determined by the condition that exponential growth has enlarged the cluster to size $\sim \lambda_{\mathrm{dB}}$.

\section{Numerical Method}
We adopt the wave–Schwarzschild method following \cite{yavetz2022} to build the FDM halo without the expensive Schrodinger-Poisson simulation.
This approach generalizes the classical Schwarzschild orbit
superposition technique to wave dark matter.
Halos are constructed as linear combinations of stationary
eigenmodes of the Schr\"odinger equation.
The amplitudes of the modes are calculated to reproduce the target density profile.
The time-dependent halo density follows from the eigenmode superposition:
\begin{equation}
\rho_{\mathrm{h}}(\mathbf{x},t) = m_a\left| \sum_{n\ell m} a_{n\ell m} \psi_{n\ell m}(\mathbf{x}) e^{-iE_{n\ell m}t/\hbar} \right|^2,
\label{eq:density}
\end{equation}
where $a_{n\ell m}$, $\psi_{n\ell m}$ and $E_{n\ell m}$ are the amplitude, eigenfunction and energy of each eigenmode labeled by $n,\ell,m$.
Interferences between modes naturally generate the desired density fluctuations and soliton random walk \cite{Li2021}.

The underlying halo density profile is chosen to be an Navarro–Frenk–White (NFW) profile superposed with a central soliton core.
The NFW component is
\begin{align}
    \rho_{\rm NFW} = \frac{\rho_0}{(r/r_s)(1+r/r_s)^2}.
\end{align}
The core density profile is described by an approximation of the soliton solution to the SP equation\cite{Schive2014}:
\begin{align}
    \rho_c =0.019\, \frac{(m_a/10^{-22}\mathrm{eV})^{-2}(r_c/\mathrm{kpc})^{-4}}{[1 + 0.091(r/r_c)^2]^8}M_\odot/\mathrm{pc}^3.
\end{align}
The core radius obeys the scaling relation
\begin{align}
    r_c = 1.6\left(\frac{m_a}{10^{-22}\mathrm{eV}}\right)^{-1}\left(\frac{M_\text{vir}}{10^9M_\odot}\right)^{-\alpha}\,\mathrm{kpc}.
\end{align}
In our simulations we adopt
\begin{align}
    M_{\rm vir}=10^9M_\odot, \quad r_s=10\,\mathrm{kpc},\quad \alpha=1/3.
\end{align}

We split the halo density into two parts $\rho=\rho_{\mathrm{base}}+\rho_{\mathrm{per}}$ where $\rho_{\mathrm{base}}(\mathbf{x}) = m_a\sum |a_{n\ell m}|^2|\psi_{n\ell m}|^2$ is the time-averaged density that sources the static potential, and $\rho_{\mathrm{per}}(\mathbf{x},t)$ contains the oscillatory interference terms with typical amplitude $\delta\rho/\rho\sim\mathcal{O}(1)$ responsible for the characteristic granular structure of wave dark matter halos.

To prevent numerical artifacts from abruptly introducing perturbations, we linearly ramp up $\rho_{\mathrm{per}}$ from $0$ to full amplitude over the first $1\,\mathrm{Gyr}$ of the simulation. The system evolves for a total of $11\,\mathrm{Gyr}$ with 110 time steps.

For each particle mass $m_a$, we generate 20 independent halo realizations with different random seeds for the eigenmode phases, allowing us to statistically characterize the variability in heating rates arising from the stochastic interference pattern.

We integrate the trajectories of stars in the stellar cluster in the time–dependent gravitational potential produced by the halo using the AGAMA \cite{agama2019} package.
The stellar cluster contains $10^4$ stars treated as zero-mass particles in the simulations.
Initially, the distribution of the stars follows the Plummer density profile
\begin{equation}
    \rho_\text{star}(r)=\frac{3M}{4\pi R_{1/2}^3}\left(1+\frac{r^2}{R^2_{1/2}}\right)^{-5/2}.
\end{equation}
Here $R_{1/2}$ is the projected (2D) half-light radius, defined as the median distance from stars to the center of mass in the xy-plane projection. We performed simulations with initial $R_{1/2} = 5,\,10,\,20\,\mathrm{pc}$. 
To assign velocities consistently with the adopted spatial distribution, we initialize the cluster by sampling AGAMA's isotropic ``QuasiSpherical'' distribution function. For the specified Plummer tracer density, AGAMA obtains the corresponding equilibrium distribution function by Eddington inversion in the static, spherically averaged potential generated by $\rho_{\mathrm{base}}$; its ``GalaxyModel'' then draws the full six-dimensional phase-space realization. The velocities are therefore sampled from an equilibrium distribution in the time-averaged halo potential rather than prescribed using circular orbits or a local Jeans approximation. Because the stars are treated as massless tracers, stellar self-gravity is consistently omitted from the construction of the initial conditions.

\section{Results}

\subsection{Evolution of Stellar Clusters}

\begin{figure*}
\includegraphics[scale=0.45]{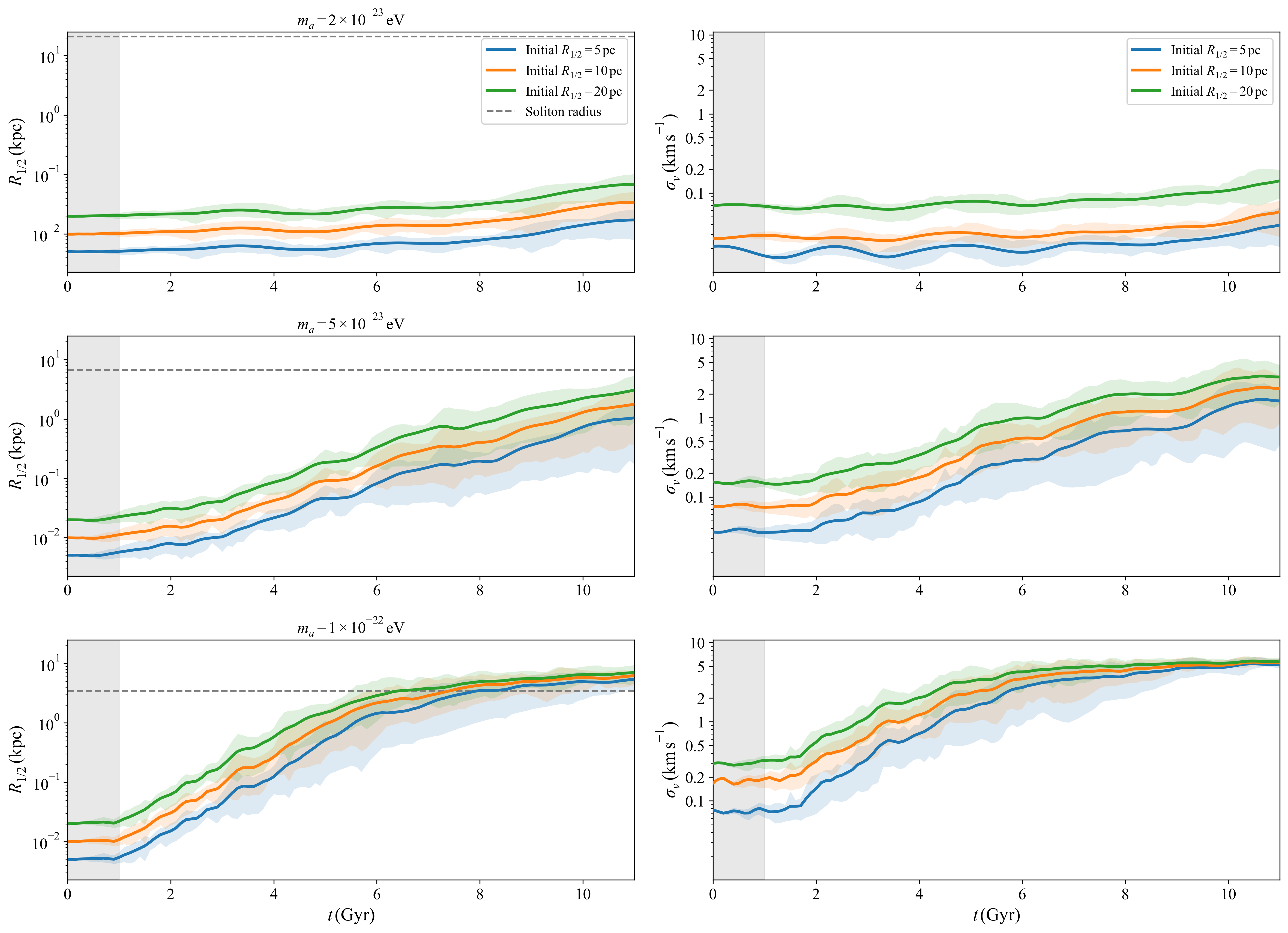}
\caption{\label{results}
Time evolution of the half-light radius $R_{1/2}$ (left) and velocity dispersion $\sigma_v$ (right) for three different FDM particle masses $m_a$. 
Solid lines are the mean value of different simulations. 
The dashed gray line marks the radius of the soliton core, which also approximates the de Broglie wavelength. 
The gray shaded region indicates the $1$~Gyr warm-up phase. 
}
\end{figure*}

Figure~\ref{results} shows the time evolution of the half-light radius $R_{1/2}$ and velocity dispersion $\sigma_v$ for three representative FDM masses: $m_a = 2\times10^{-23}$ eV, $5\times10^{-23}$ eV and $10^{-22}$ eV. Solid lines represent averages over simulations with different initial $R_{1/2}$ ($5, 10, 20$~pc), with shaded regions indicating the range of variation. The gray shaded region ($t<1$ Gyr) indicates the warm-up phase where density fluctuations are slowly turned on. The dashed horizontal lines mark the soliton core radius, which also approximates $\lambda_{\mathrm{dB}}$.

\begin{figure*}
\includegraphics[scale=0.52]{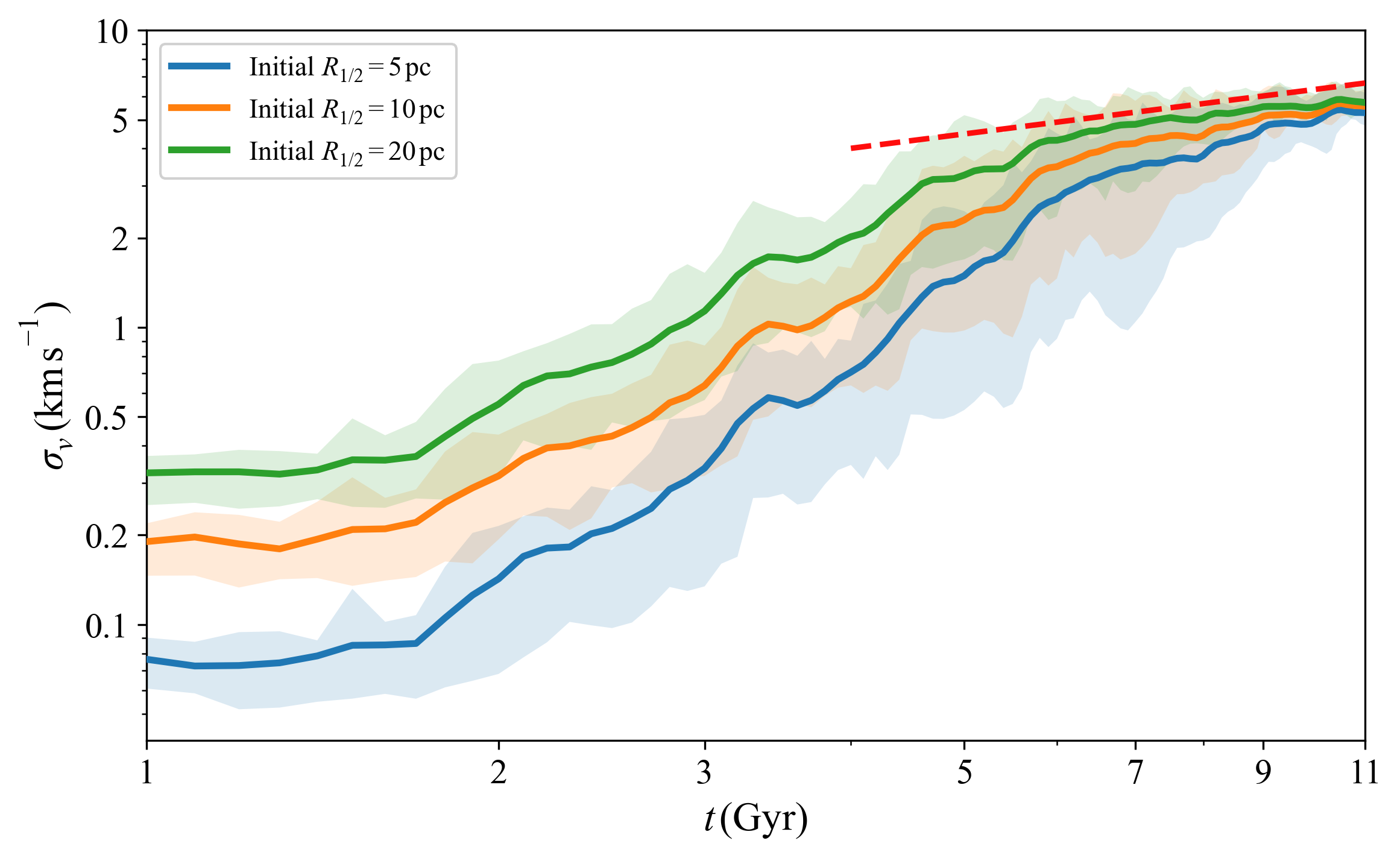}
\caption{\label{loglog} Log-log plot for the evolution of $\sigma_v$ with \(m_a = 1.0 \times 10^{-22} \, \mathrm{eV}\). Clusters enter the diffusion heating stage after its size reached the soliton radius. The red dashed line represents $\sigma_v\propto \sqrt{t}$ trend as shown in Equation~\ref{eq:heating}.}
\end{figure*}

The evolution proceeds in three distinct stages:
\begin{enumerate}
    \item Initial phase ($t<1$ Gyr): During the warm-up, both $R_{1/2}$ and $\sigma_v$ remain nearly constant, serving as effective initial conditions.
    \item Exponential growth phase: After the full fluctuating density field is activated, both quantities grow exponentially.
    Runs with different initial $R_{1/2}$ follow the same exponential trend; the cluster size at any given time is proportional to the initial $R_{1/2}$. This exponential growth is the hallmark of tidal heating.
    \item Diffusive phase: Once $R_{1/2}$ reaches the soliton core radius (dashed lines), exponential growth terminates and heating transitions to a diffusive regime. Figure \ref{loglog} shows the velocity dispersion evolution for $m_a = 10^{-22}$ eV on a log-log scale. At late times, $\sigma_v(t)$ follows $\sigma_v \propto \sqrt{t}$ (red dashed line), exactly as predicted by Equation~\ref{eq:heating}. Clusters with different initial sizes eventually converge to the same $\sigma_v$ as they become virialized with the FDM halo.
\end{enumerate}

The dependence on $m_a$ is striking. For $m_a = 2\times10^{-23}$ eV (top row of Figure 1), the exponential growth rate is so slow that over 10 Gyr the cluster expands only modestly. For $m_a = 5\times10^{-23}$ eV (middle row), the cluster is still in the exponential growth phase at the end of the simulation and has not yet entered diffusive heating. For $m_a = 10^{-22}$ eV (bottom row), the cluster reaches $\lambda_{\mathrm{dB}}$ at around 7 Gyr and subsequently enters the diffusive regime.

\begin{figure*}
\includegraphics[scale=0.4]{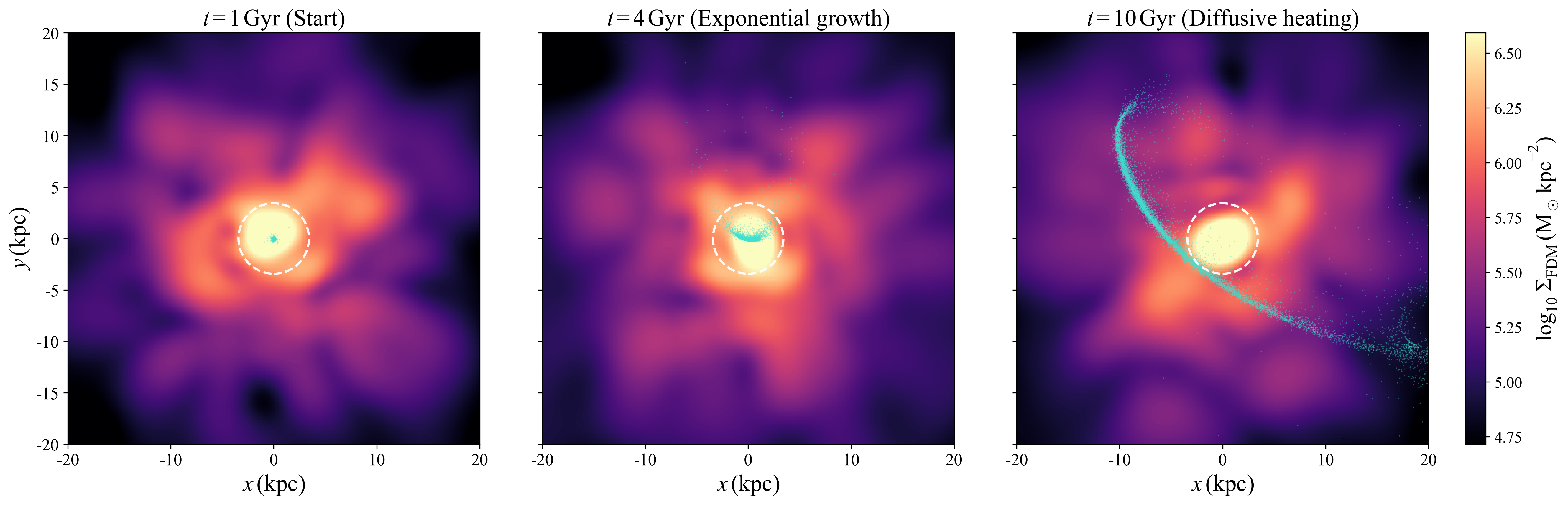}
\caption{\label{cluster evolution}
Snapshots of one simulation with \(m_a = 1.0 \times 10^{-22} \, \mathrm{eV}\) at $1\,\mathrm{Gyr}$, $4\,\mathrm{Gyr}$ and $10\,\mathrm{Gyr}$. Cyan points are stellar particles overlaid on the projected FDM surface density map. The white dashed circle marks the soliton-core radius.
}
\end{figure*}

Figure~\ref{cluster evolution} shows snapshots of one simulation with $m_a = 10^{-22}$ eV and initial $R_{1/2}=20$ pc. At $t=1$ Gyr (left panel), the cluster is compact and roughly spherical. At $t=4$ Gyr (middle panel), during the exponential expansion phase, the cluster is stretched and becomes non-spherical, with its size comparable to the soliton core. At $t=10$ Gyr (right panel), after exponential growth has terminated, the cluster has been deformed into a stream much larger than the soliton core, and heating is now diffusive (as seen in Figure~\ref{loglog}).

\begin{figure*}
\includegraphics[scale=0.52]{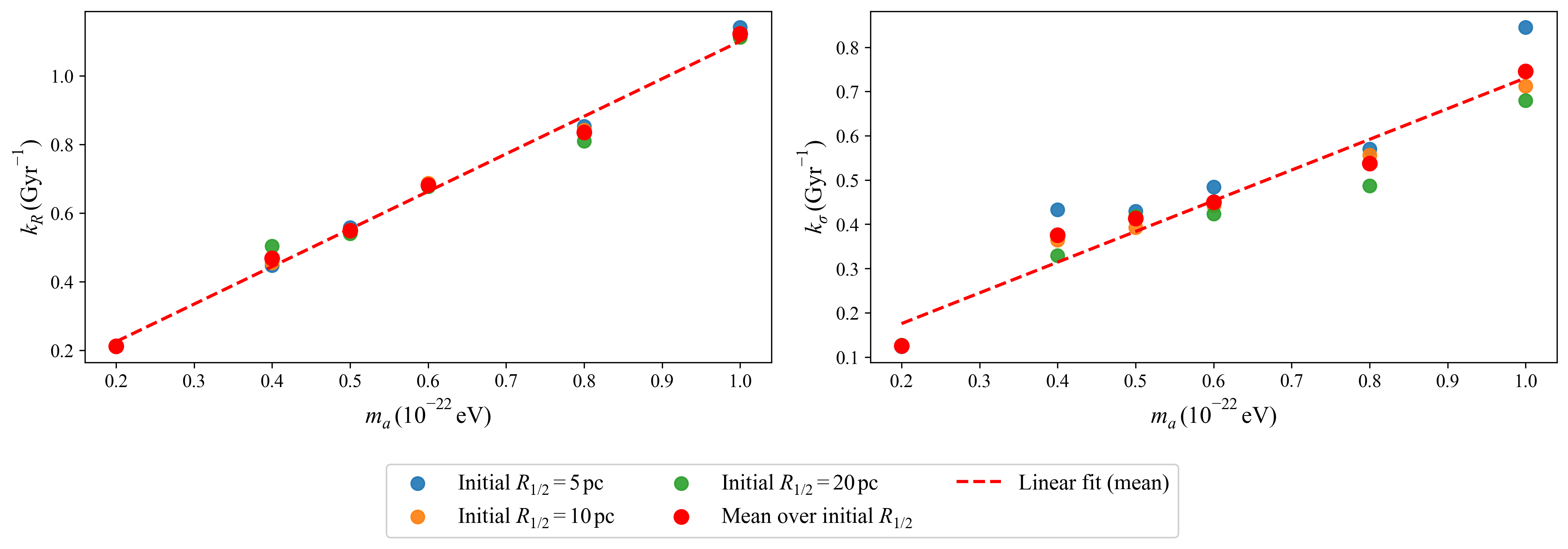}
\caption{\label{k_vs_ma} 
Dependence of the exponential growth rates $k_R$ (left) and $k_\sigma$ (right) on the FDM particle mass $m_a$. 
Fitted red lines have slope $1.09\,\mathrm{Gyr}^{-1}(10^{-22}\,\mathrm{eV})^{-1}$ and $0.69\,\mathrm{Gyr}^{-1}(10^{-22}\,\mathrm{eV})^{-1}$ separately.
}
\end{figure*}

Figure~\ref{k_vs_ma} quantifies the exponential growth rates $k_R$ and $k_\sigma$ as functions of $m_a$. Blue, orange, and green points correspond to different initial $R_{1/2}$, while red points are the mean for each $m_a$. The red dashed lines show linear fits: $k_R \propto m_a$ with slope $1.09$ Gyr$^{-1}(10^{-22}$ eV$)^{-1}$, and $k_\sigma \propto m_a$ with slope $0.69$ Gyr$^{-1}(10^{-22}$ eV$)^{-1}$.
This linear scaling is precisely the behavior predicted by tidal heating.
Using Equation~\ref{growthrate} with our satellite halo parameters, we can calculate
\begin{equation}
    k_R\sim 1.5\left(\frac{m_a}{10^{-22}\mathrm{eV}}\right)\,
    \mathrm{Gyr}^{-1}
\end{equation}
which is in good agreement with our simulation results.

\begin{figure*}
\centering
\includegraphics[width=0.88\textwidth]{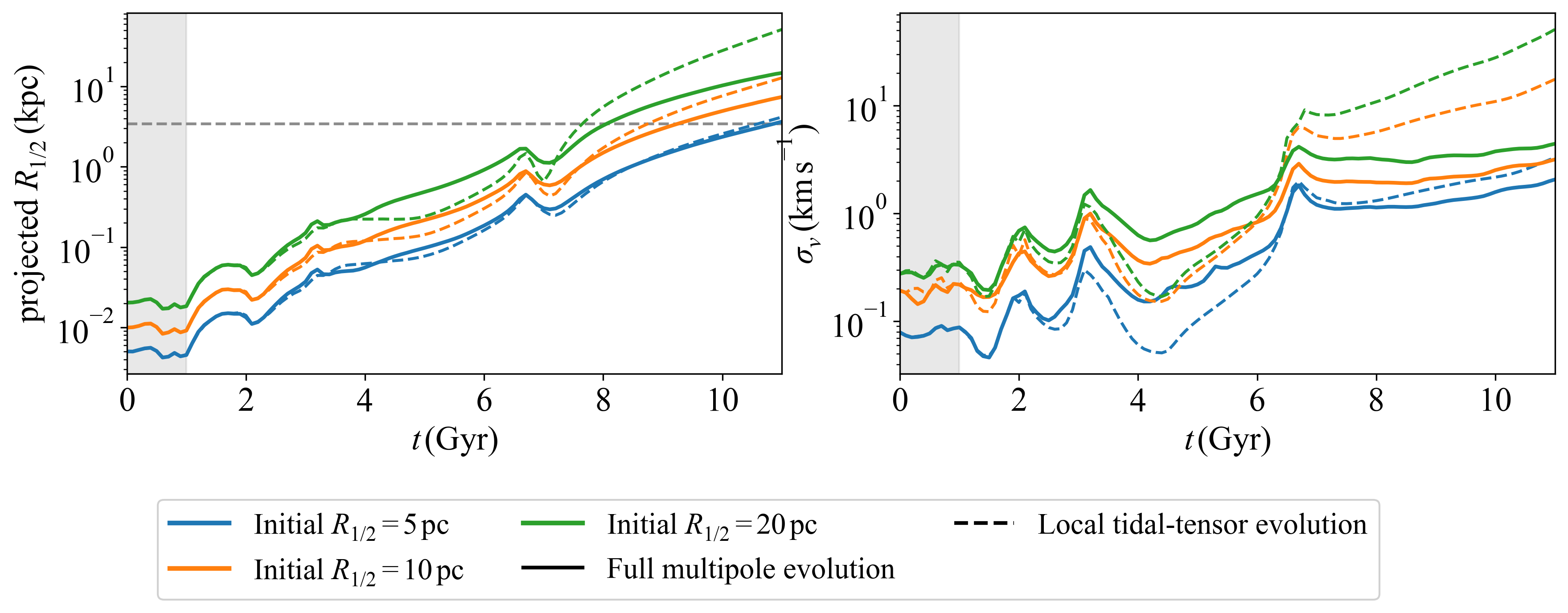}
\caption{Comparison of the stellar cluster evolution from the full simulation (solid lines) and the local tidal equation (dashed lines) for one realization with $m_a=1.0\times10^{-22}\,\mathrm{eV}$. The full simulation uses the same numerical configuration as the main simulations. The tidal equation reproduces the exponential growth phase of the cluster size until it reaches the soliton core radius, supporting tidal heating as the dominant mechanism in this regime. The local approximation does not capture the later diffusive phase.}
\label{fig:tidal_equation}
\end{figure*}

{
We further test the physical interpretation of the exponential-growth phase using one realization with $m_a=1.0\times10^{-22}\,\mathrm{eV}$ and the same numerical configuration as in the main simulations. We integrate the local tidal equation
\begin{align*}
    \ddot{\xi}_i(t)=-T_{ij}(t)\xi_j(t),
\end{align*}
where $\boldsymbol{\xi}=\mathbf{x}-\mathbf{R}_{\rm cm}$ and $T_{ij}=\partial_i\partial_j\Phi$ is evaluated along the measured center-of-mass trajectory. Figure~\ref{fig:tidal_equation} shows that this local model reproduces the exponential expansion until the cluster becomes comparable in size to the soliton core. Its failure to reproduce the subsequent diffusive phase is expected because the local linearization is no longer adequate on those scales. This agreement supports the interpretation that the exponential phase is a physical response to the time-dependent tidal field of the FDM halo.

Additional control experiments and numerical convergence tests are presented in Appendix~\ref{app:controls}.
}

\begin{figure*}[t]
\includegraphics[scale=0.52]{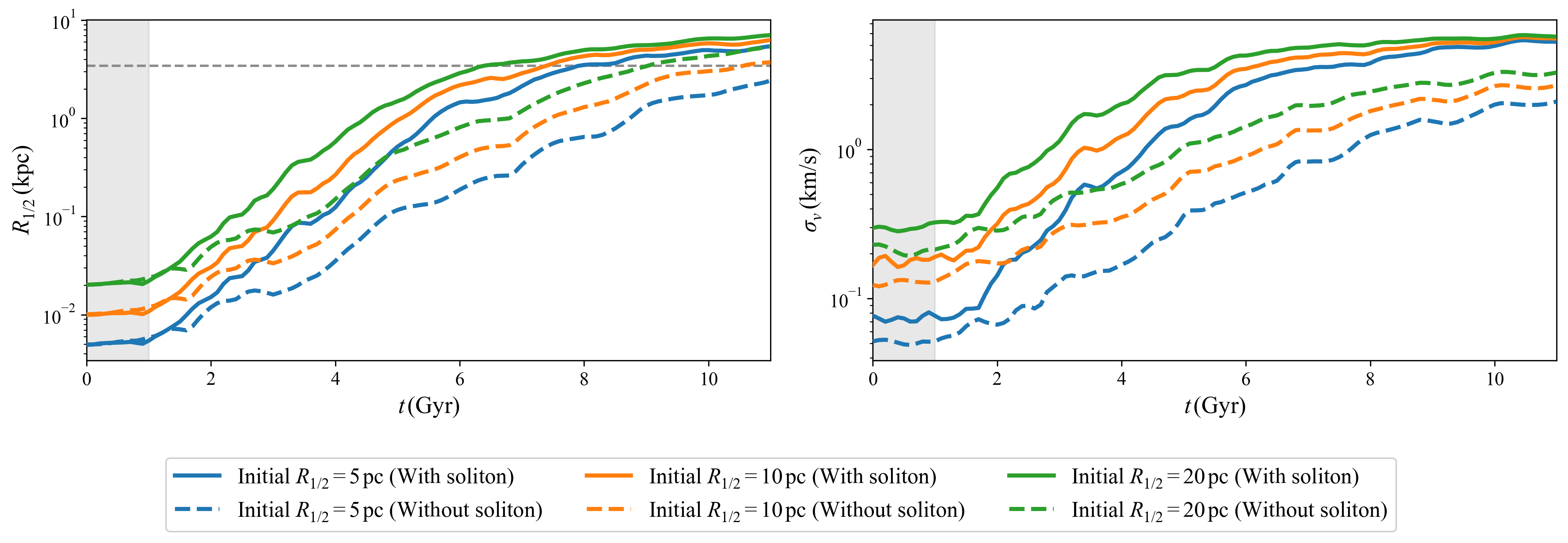}
\caption{\label{heating without soliton}
Evolution of $R_{1/2}$ and $\sigma_v$ for a halo with (solid lines) and without (dashed lines) a soliton core for $m_a = 1.0 \times 10^{-22} \, \mathrm{eV}$. 
}
\end{figure*}

\begin{figure*}[t]
\includegraphics[scale=0.52]{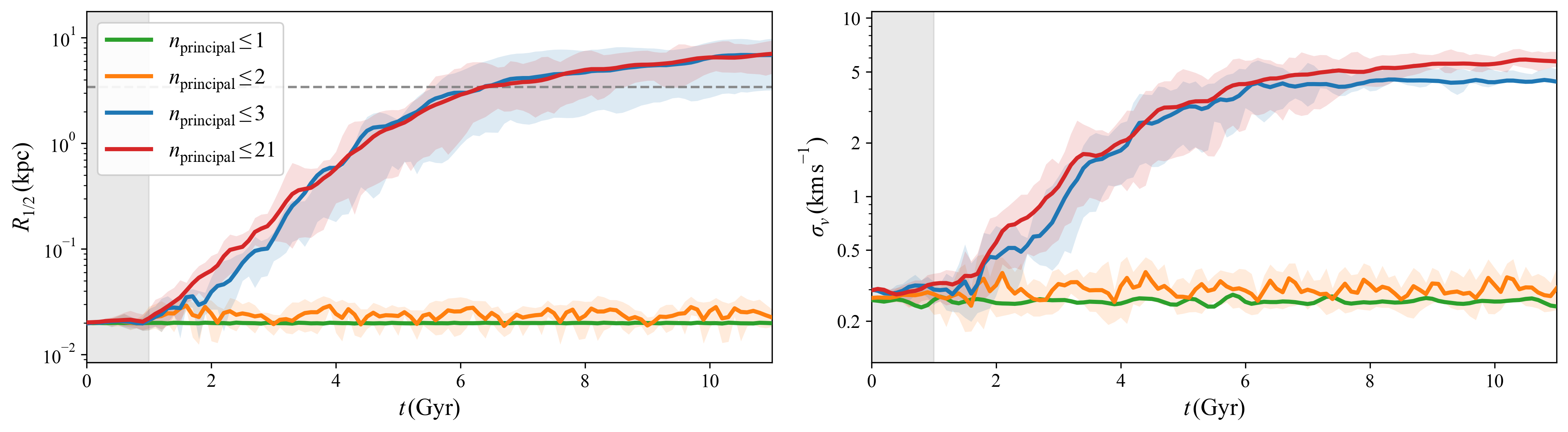}
\caption{\label{tidal}
Dependence of stellar cluster heating on the maximum principal quantum number \(n_{\mathrm{principal}}\) for \(m_a = 1.0 \times 10^{-22} \, \mathrm{eV}\). The full halo model (red) includes modes up to \(n_{\mathrm{principal}} \leq 21\), capturing both granular interference and soliton random walk. 
}
\end{figure*}

\FloatBarrier
\subsection{Effects of the Central Soliton}
The soliton core at the halo center exhibits a random walk with amplitude comparable to its core radius \cite{Schive2020,Li2021}.
This motion could enhance potential fluctuations and thus cluster heating. 
In the previous simulations, we employ a standard FDM radius-halo mass relation $M_c\propto M_{\rm vir}^{1/3}$ \cite{Schive2014b}.
However, different soliton mass can also yield stable halo configurations \cite{yavetz2022}.
To isolate this effect, we performed control simulations without a soliton (NFW only) for $m_a = 1.0\times10^{-22}$ eV. In these halos, the time-averaged density has a smaller ground-state amplitude, and soliton random walk is suppressed.

Figure~\ref{heating without soliton} compares results with and without a soliton (solid and dashed lines, respectively). Exponential growth still occurs without the soliton, but the growth rate is smaller. Removing the soliton reduces the central potential $\Phi_0$, leading to slower heating. At the end of the simulation, the velocity dispersion in the no-soliton case remains lower than in the soliton case. Soliton random walks therefore provide an additional, significant heating channel.

\subsection{Heating from a Tidally Stripped Halo}
A major objection to \cite{dalal2022} is that the host Milky Way halo may tidally strip the satellite FDM halo, removing the granular structure and leaving only the central soliton, thereby suppressing heating \cite{Schive2020,Yang2025}.
To investigate this, we approximate the tidally stripped satellite halos by limiting the maximum principal quantum number $n_{\mathrm{principal}}\equiv n+\ell+1$ in our halo construction for $m_a = 1.0\times10^{-22}$ eV. Our full halo in the previous sections includes modes up to $n_{\mathrm{principal}}\leq 21$. Reducing $n_{\mathrm{principal}}$ progressively removes higher-order modes, mimicking tidal stripping. The $n_{\mathrm{principal}}=1$ case retains only the spherically symmetric, non-evolving $n=0$, $\ell=0$ mode, corresponding to a fully stripped halo with only a static soliton \cite{Schive2020}.

Figure~\ref{tidal} shows the results with different maximal $n_{\mathrm{principal}}$.
For $n_{\mathrm{principal}}=1$ (fully stripped), heating is completely suppressed: the cluster size and velocity dispersion remain nearly constant. For $n_{\mathrm{principal}}\leq 2$ (including azimuthal modes $m=0,\pm1$), density fluctuations are still mainly radial, and no significant heating occurs. When $n_{\mathrm{principal}}\leq 3$ are included, the heating approaches the full-halo case. Thus, tidal stripping of higher-order eigenmodes, i.e., removal of the wave-induced granularity, can dramatically reduce heating efficiency. 
This finding suggests that the heating of UFDs in FDM models depends sensitively on their orbital history and the degree of tidal stripping.

\long\def\controlsappendix{
\section{Controls and Convergence Tests}
\label{app:controls}
\begin{table*}[t]
    {
    \centering
    \caption{Physical and numerical parameters used for the main simulations. The final column lists the values used in the corresponding convergence tests; a dash denotes a parameter that was not varied in these tests.}
    \label{tab:convergence}
    \begin{ruledtabular}
    \begin{tabular}{lccc}
        Quantity & Symbol or option & Main simulations & Values tested \\
        \hline
        FDM particle mass & $m_a\ (10^{-22}\,\mathrm{eV})$ & $0.2,\ 0.4,\ 0.5,\ 0.6,\ 0.8,\ 1.0$ & --- \\
        Halo mass & $M$ & $10^9\,M_\odot$ & --- \\
        Initial projected half-light radius & $R_{1/2}(0)$ & $5,\ 10,\ 20\,\mathrm{pc}$ & --- \\
        Number of realizations & $N_{\mathrm{seed}}$ & $20$ & --- \\
        Evolution interval & $t$ & $0$--$11\,\mathrm{Gyr}$ & --- \\
        Initial radial-profile samples & $N_{\mathrm{setup}}$ & $512$ & --- \\
        Multipole radial grid points & $N_r$ & $25$ & $25,\ 50,\ 100$ \\
        Multipole angular degree & $L_{\max}$ & $20$ & $8,\ 12,\ 16,\ 20,\ 23$ \\
        Multipole azimuthal limit (default) & $M_{\max}$ & $20$ & $M_{\max}=L_{\max}$ in each case \\
        Maximum principal quantum number & $n_{\mathrm{principal}}$ & $21$ & $5,\ 9,\ 13,\ 17,\ 21$ \\
        Potential-snapshot cadence & $\Delta t$ & $0.1\,\mathrm{Gyr}$ & $0.05,\ 0.1,\ 0.2\,\mathrm{Gyr}$ \\
        Temporal interpolation & \texttt{linearInterp} & linear & linear, cubic spline \\
        AGAMA trajectory samples & \texttt{trajsize} & $1000$ & $200,\ 500,\ 1000,\ 2000,\ 4000$ \\
        ODE relative accuracy & \texttt{accuracy} & $10^{-8}$ & $10^{-2},\ 10^{-4},\ 10^{-8},\ 10^{-10},\ 10^{-12}$ \\
    \end{tabular}
    \end{ruledtabular}
    }
\end{table*}
{
We performed a suite of controlled experiments and convergence tests to establish that the measured cluster heating is driven by the time dependence of the FDM potential and is not produced by the numerical representation or orbit integration. Unless stated otherwise, these tests use a single halo realization with $m_a=1.0\times10^{-22}\,\mathrm{eV}$ and retain the same stellar initial conditions when comparing numerical configurations. The fiducial configuration is shown by solid curves in each figure, and the parameters varied in the tests are summarized in Table~\ref{tab:convergence}.

We first isolate the role of temporal fluctuations by comparing the fiducial evolving halo with two time-independent controls. In the static-base control, the fluctuating component is removed by setting $\rho_{\rm per}=0$. In the frozen control, the fully developed perturbation is held fixed at $\rho_{\rm per}(\mathbf{x},t)=\rho_{\rm per}(\mathbf{x},1\,\mathrm{Gyr})$ throughout the subsequent evolution. Figure~\ref{static comparison} shows that neither control produces heating comparable to that in the evolving halo. Because the stellar realization is unchanged, this comparison identifies the temporal variation of the FDM potential, rather than the presence of spatial granularity alone, as the source of the secular increase in $R_{1/2}$ and $\sigma_v$.

We next test the temporal sampling and interpolation of the evolving potential. Independent density and potential sequences were generated with snapshot cadences of $0.05$, $0.1$, and $0.2\,\mathrm{Gyr}$ and evolved over the same $11\,\mathrm{Gyr}$ interval. The agreement among the resulting cluster histories in the left panel of Figure~\ref{temporal interpolation} demonstrates that the fiducial cadence of $0.1\,\mathrm{Gyr}$ resolves the relevant potential variability. The right panel compares linear and cubic-spline interpolation between snapshots and shows that the inferred evolution is insensitive to the interpolation scheme. These tests concern the sampling and interpolation of the external potential and are distinct from the adaptive time stepping of the orbit integrator considered below.

The finite eigenmode representation is tested separately from the tidal-stripping experiment in Figure~\ref{tidal}. Using the same $n_{\mathrm{principal}}$ cutoff defined for that experiment, we reconstruct a single halo realization with fixed mode amplitudes and phases and successively retain modes satisfying $n_{\mathrm{principal}}\leq5,9,13,17,$ and 21. For each cutoff, the full time-dependent potential sequence is regenerated and the same stellar phase-space realization is evolved with identical potential-expansion and orbit-integration settings. Thus, differences among these runs measure only the numerical sensitivity to the eigenmode cutoff. The sequences in Figure~\ref{eigenmode convergence} approach the fiducial cutoff of 21 as higher-order modes are added. At the final time, the cutoff-17 and cutoff-21 calculations differ by $4$--$14\%$ in $R_{1/2}$ and $15$--$20\%$ in $\sigma_v$. The agreement between these two highest cutoffs indicates that the adopted cutoff of 21 is approximately converged, with the remaining differences providing an estimate of the residual eigenmode-truncation uncertainty.

We assess the spatial representation of the potential using the AGAMA multipole expansion. To distinguish this representation from the FDM eigenmode basis, we reserve lowercase $(n,\ell,m)$ for the Schr\"odinger eigenmodes in Eq.~\eqref{eq:density} and use uppercase $(L,M)$ for the angular multipole indices, such that $\Phi(r,\theta,\phi)=\sum_{L=0}^{L_{\max}}\sum_{M=-L}^{L}\Phi_{LM}(r)Y_{LM}(\theta,\phi)$. The separate quantity $N_r$ is the number of radial grid nodes used to tabulate $\Phi_{LM}(r)$ and is not a mode index analogous to $n$. Because \texttt{mmax} is not specified, AGAMA uses its default $M_{\max}=L_{\max}$ and retains every azimuthal order $-L\leq M\leq L$ at each included $L$. We test this representation by varying $N_r$ and $L_{\max}$. Figure~\ref{resolution} shows that the principal features of the cluster evolution are stable across the tested spatial resolutions. As a direct force-level check, Figure~\ref{force accuracy} compares the accelerations evaluated at the evolved particle positions using the production potential, with $N_r=25$ and $L_{\max}=20$, and a higher-resolution representation, with $N_r=100$ and $L_{\max}=23$, constructed from the same density snapshots. This comparison tests the actual accelerations entering the tracer integrations and indicates whether spatial discretization errors could account for the measured heating. The finite expansion may smooth radial features not resolved by $N_r$ and omits angular structure only for $L>L_{\max}$, rather than removing selected $M$ components at retained values of $L$.

The orbit calculation is tested independently by varying both the number of stored trajectory samples and the accuracy tolerance of AGAMA's adaptive integrator. As shown in Figure~\ref{agama_convergence}, changing \texttt{trajsize} from 200 to 4000 and the relative accuracy from $10^{-2}$ to $10^{-12}$ leaves the physical evolution consistent with the fiducial choices, \texttt{trajsize}$=1000$ and \texttt{accuracy}$=10^{-8}$. The former controls the sampling of the returned trajectories, whereas the latter controls the internal adaptive integration and therefore directly probes accumulated orbit-integration error.

Taken together, the control experiments and convergence tests show that the measured heating is not an artifact of the potential representation or orbit integration, while quantifying the remaining sensitivity to the finite eigenmode basis.
}

\begin{figure*}
\centering
\includegraphics[width=0.56\textwidth]{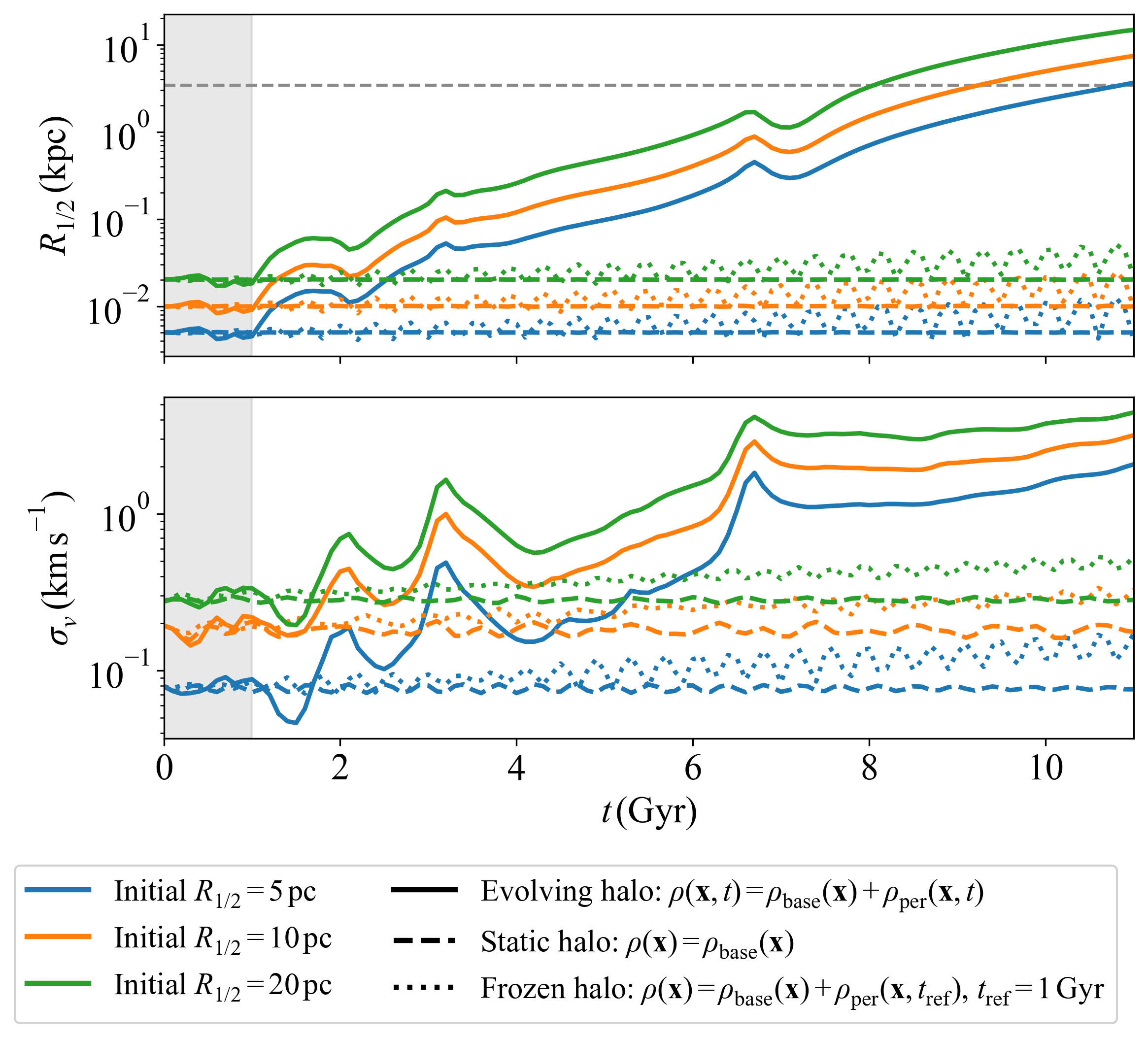}
{\caption[Time-dependence control tests.]{Comparison of stellar-cluster evolution in the fiducial evolving halo, the static base halo, and a frozen fully ramped halo. The absence of comparable heating in the time-independent controls identifies temporal FDM fluctuations as the source of the heating.}}
\label{static comparison}
\end{figure*}

\begin{figure*}
\centering
\includegraphics[width=0.485\textwidth]{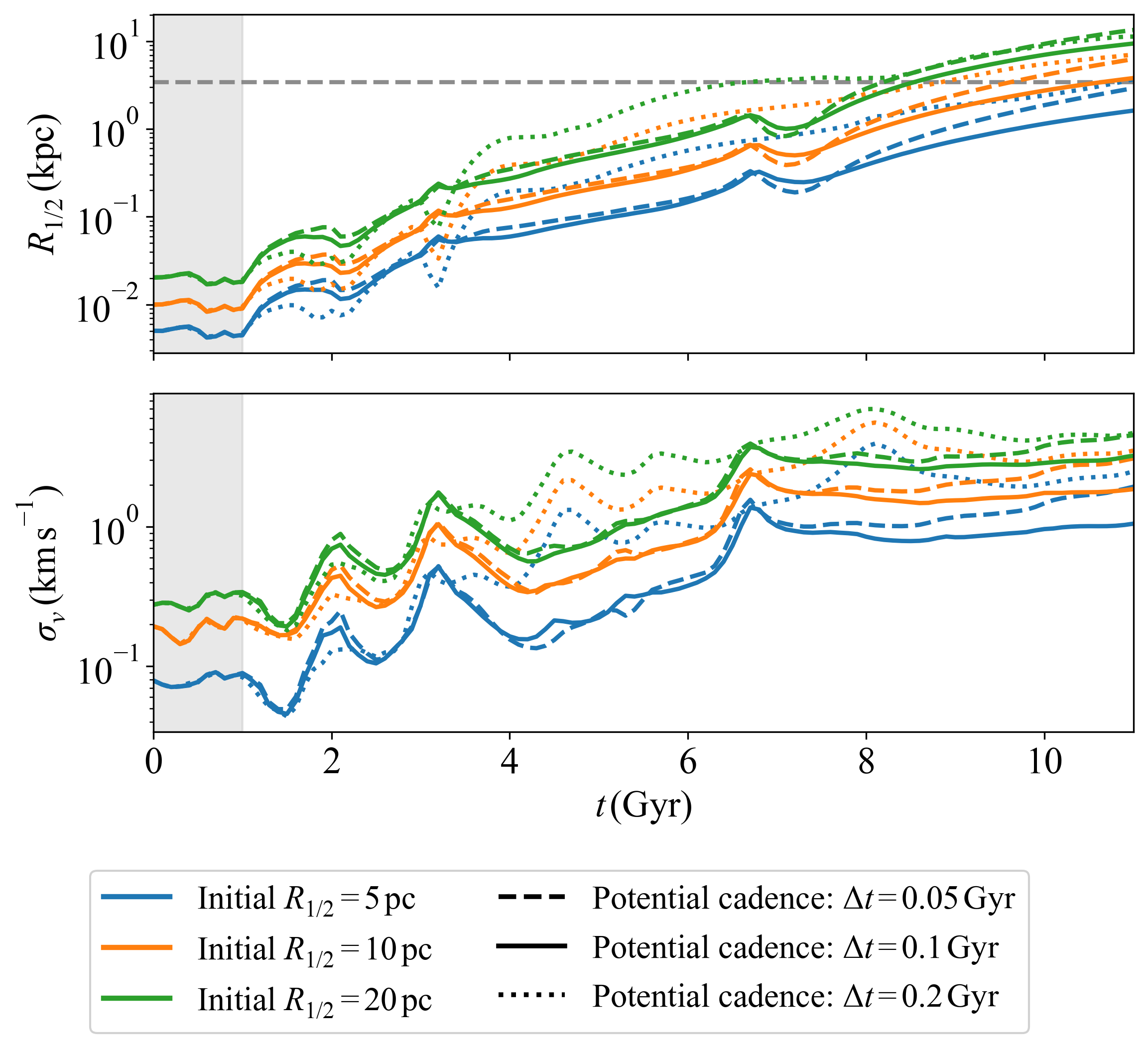}\hfill
\includegraphics[width=0.485\textwidth]{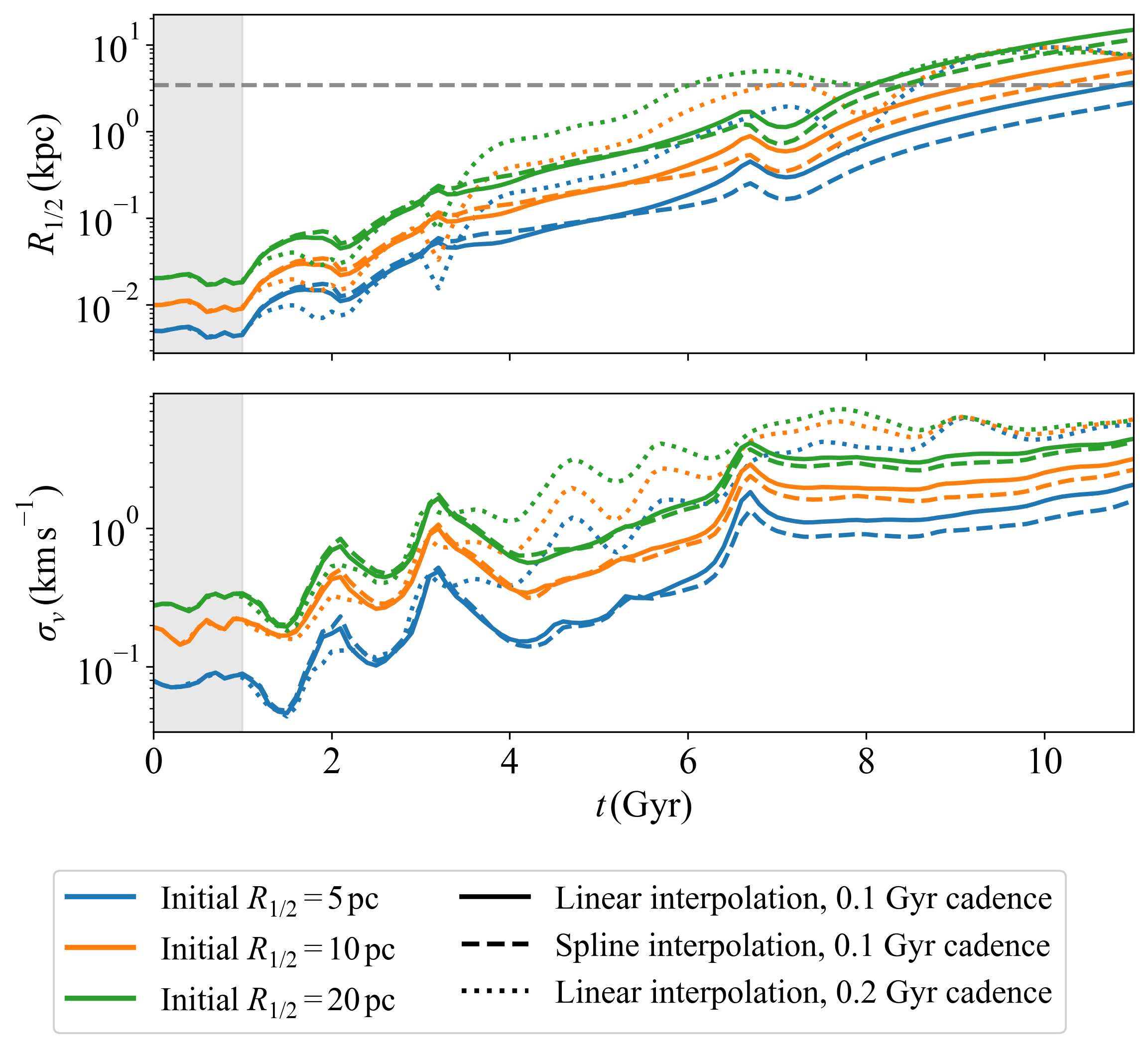}
{\caption[Temporal-convergence tests.]{Temporal-convergence tests for the stellar-cluster evolution. Left: independently regenerated potential sequences with snapshot cadences of $0.05$, $0.1$, and $0.2\,\mathrm{Gyr}$. Right: comparison of linear and cubic-spline interpolation at the fiducial $0.1\,\mathrm{Gyr}$ cadence, together with a coarser-cadence case. Colors denote the initial projected half-light radius, and line styles denote the temporal treatment.}}
\label{halo temporal convergence}
\label{temporal interpolation}
\end{figure*}

\begin{figure*}
\centering
\includegraphics[width=0.56\textwidth]{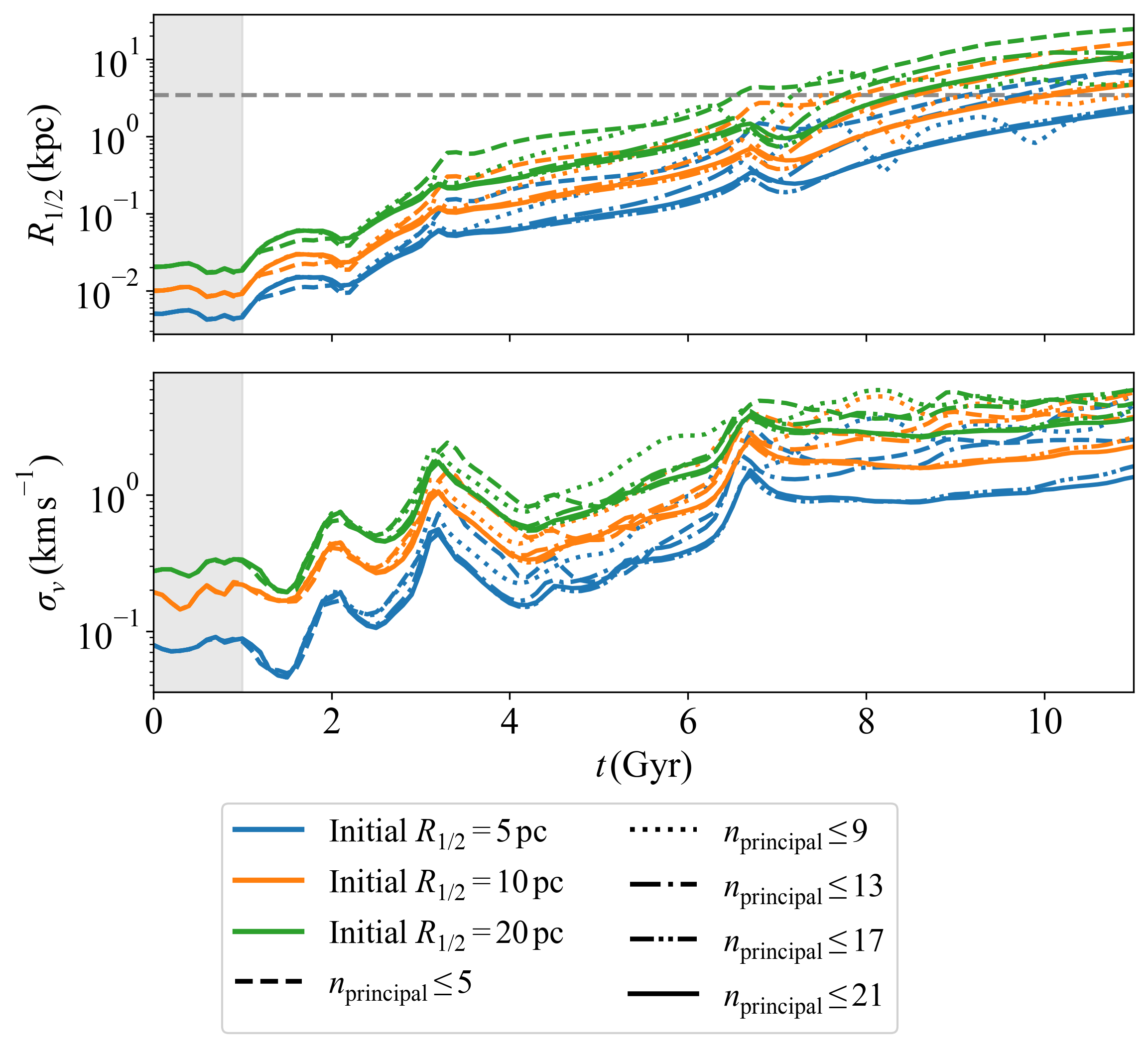}
{\caption{Same-seed eigenmode-cutoff test. The halo phases, mode amplitudes, potential representation, stellar initial conditions, and orbit settings are fixed while the maximum principal quantum number is varied. The main-simulation value $n_{\mathrm{principal}}\leq21$ is shown by solid lines. The cutoff-17 and cutoff-21 results differ at the end of the simulations by $4$--$14\%$ in $R_{1/2}$ and $15$--$20\%$ in $\sigma_v$, quantifying the residual sensitivity to the mode cutoff.}}
\label{eigenmode convergence}
\end{figure*}

\begin{figure*}
\centering
\includegraphics[width=0.485\textwidth,height=0.438\textwidth,keepaspectratio]{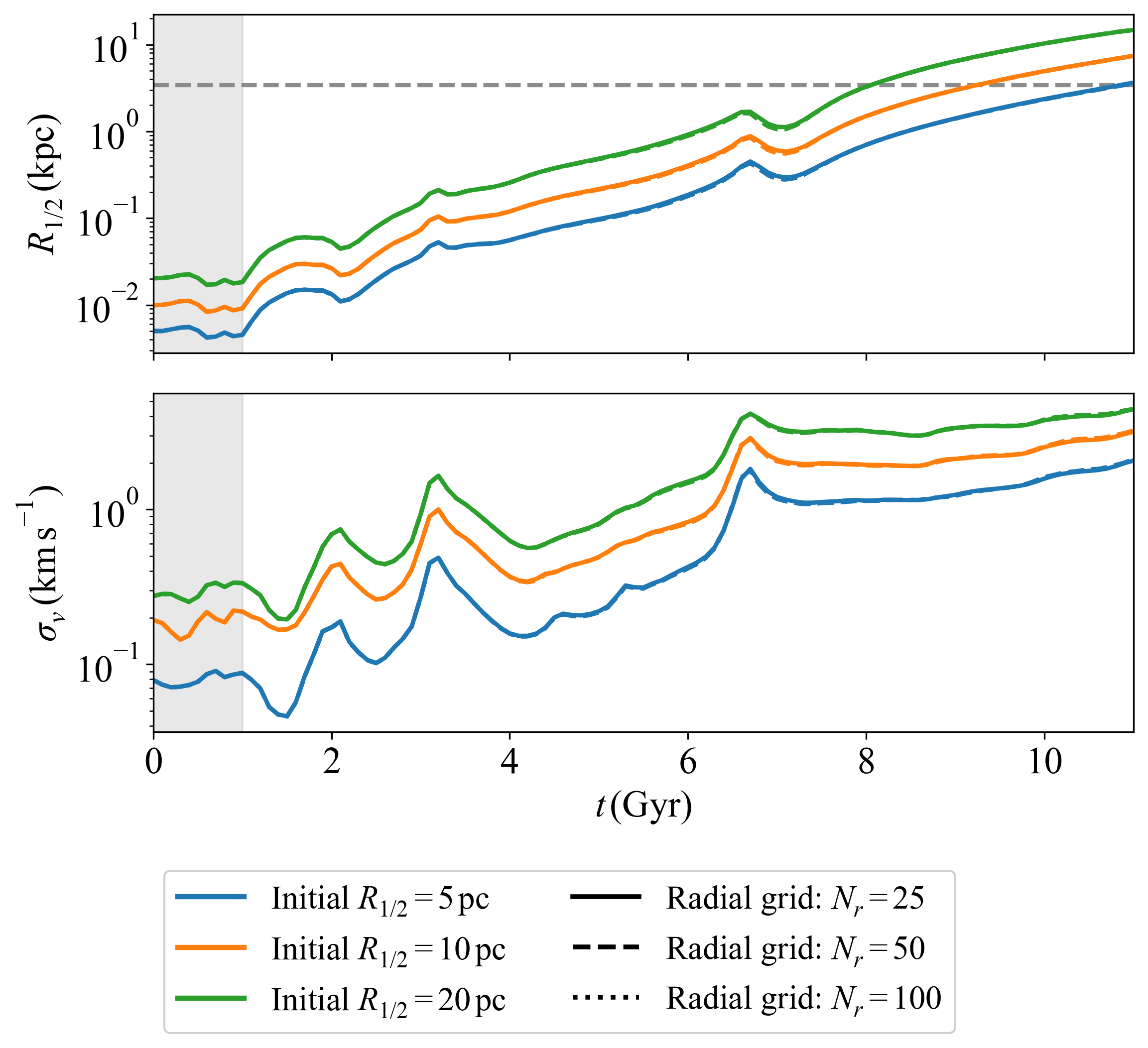}\hfill
\includegraphics[width=0.485\textwidth,height=0.438\textwidth,keepaspectratio]{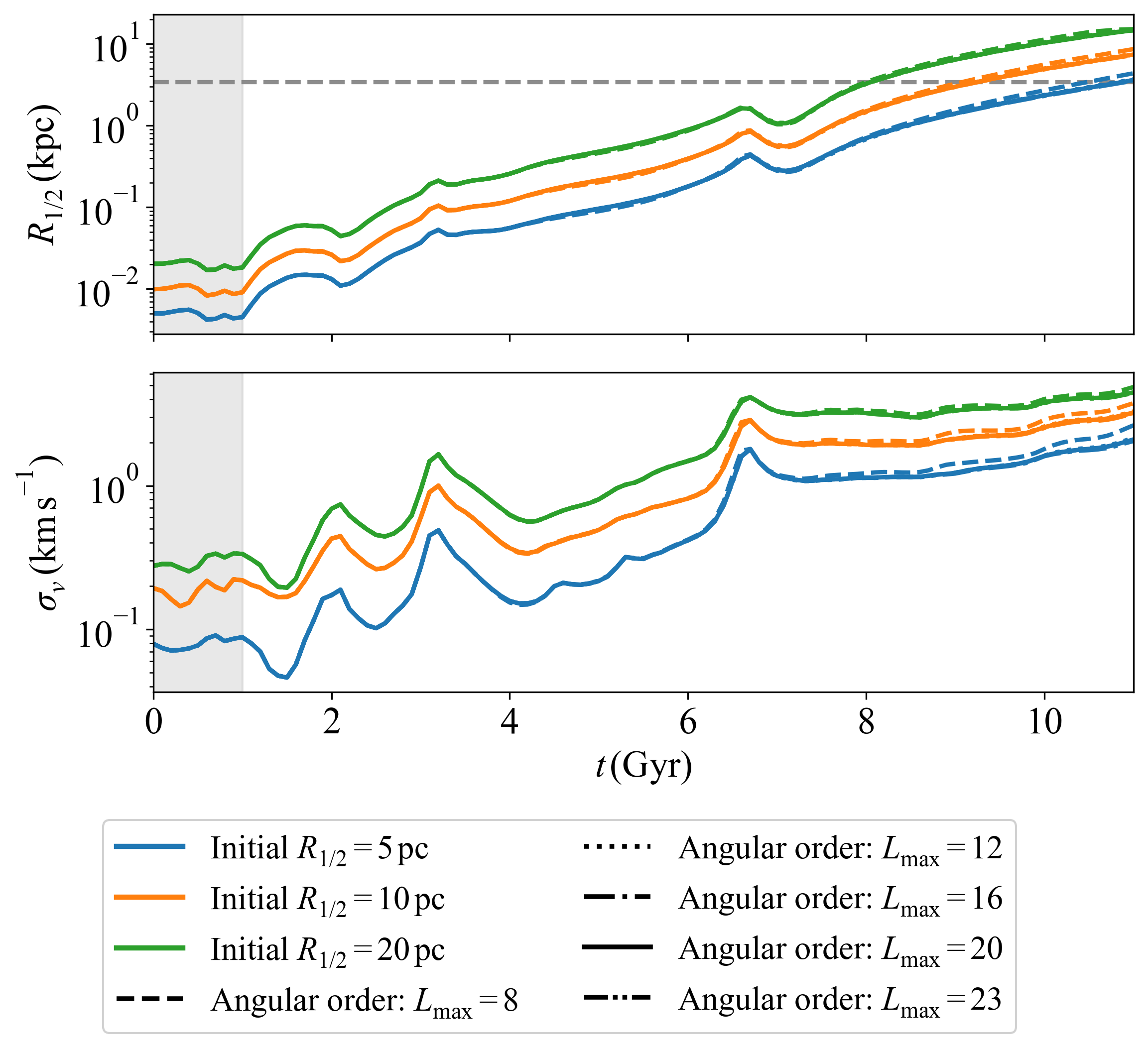}
{\caption{Spatial convergence of the stellar-cluster evolution with respect to the radial grid size $N_r$ (left) and maximum angular degree $L_{\max}$ (right) of the AGAMA multipole expansion. Colors denote the initial projected half-light radius and line styles denote the numerical resolution. The main-simulation values, $N_r=25$ and $L_{\max}=20$, are shown by solid lines.}}
\label{resolution}
\end{figure*}

\begin{figure*}
\centering
\includegraphics[width=0.485\textwidth,height=0.438\textwidth,keepaspectratio]{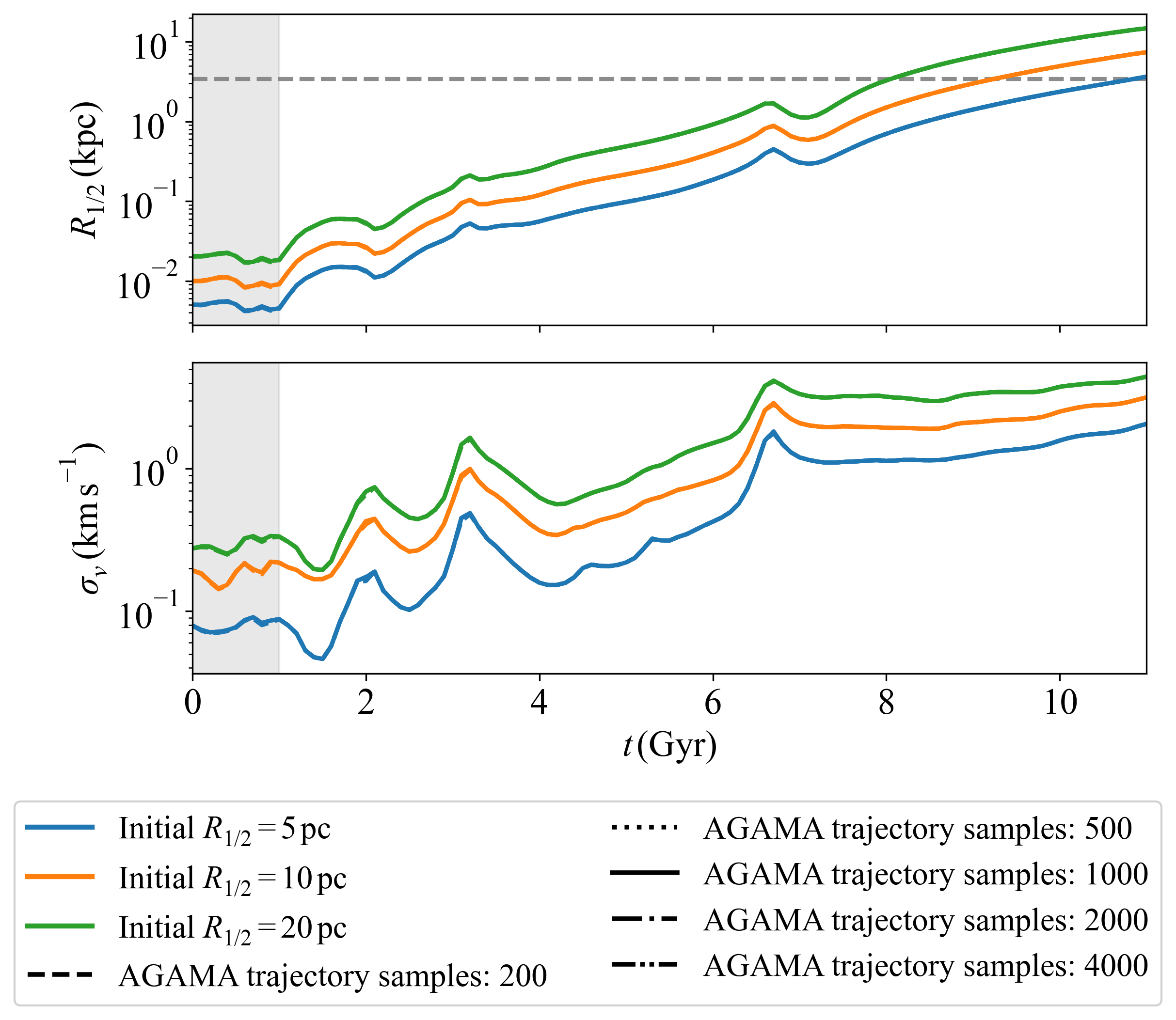}\hfill
\includegraphics[width=0.485\textwidth,height=0.438\textwidth,keepaspectratio]{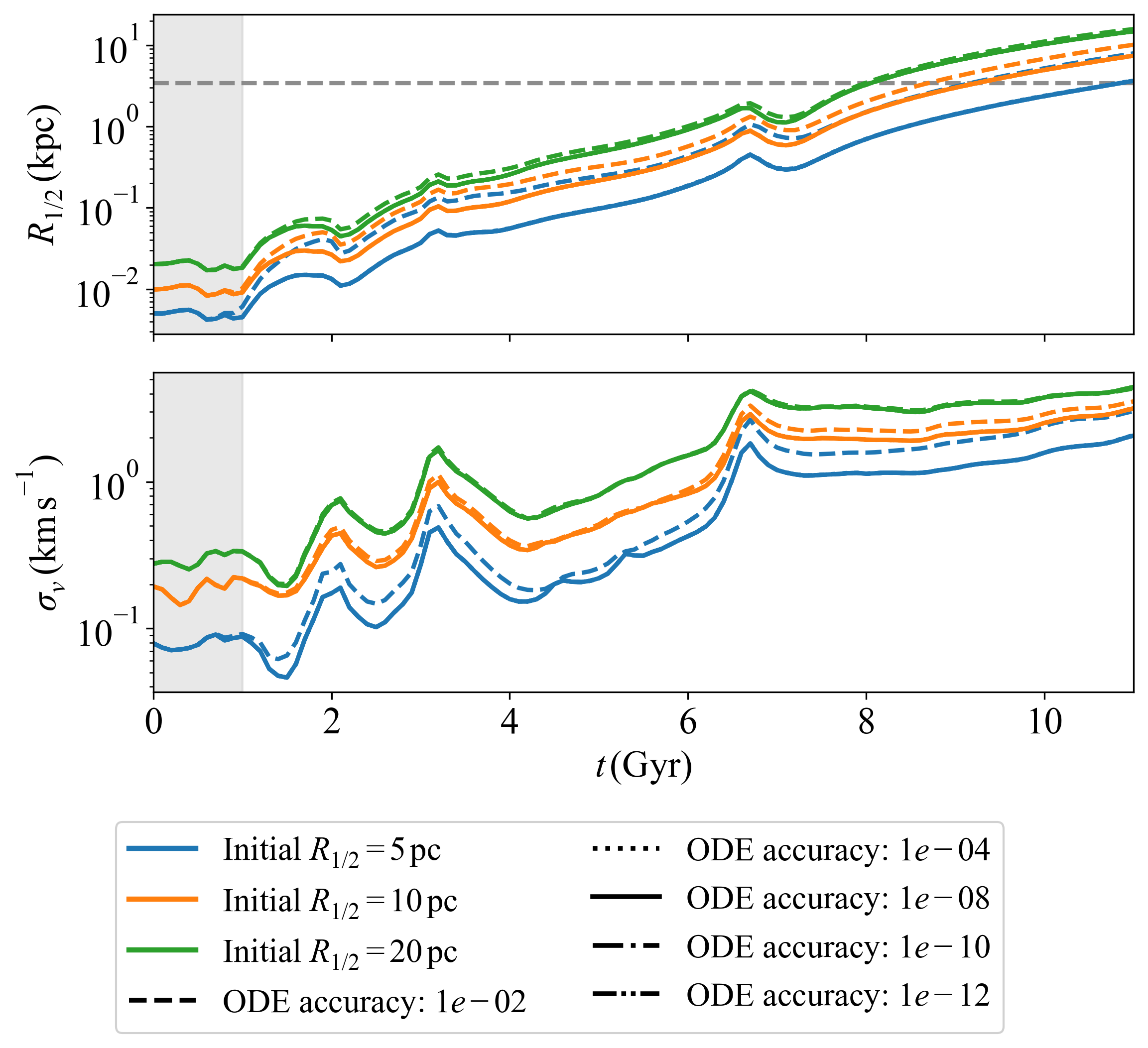}
{\caption{Convergence of the stellar-cluster evolution with respect to AGAMA's trajectory output sampling (left) and adaptive integration accuracy (right). Colors denote the initial projected half-light radius and line styles denote the numerical setting. The main-simulation values, \texttt{trajsize}$=1000$ and \texttt{accuracy}$=10^{-8}$, are shown by solid lines.}}
\label{agama_convergence}
\end{figure*}

\begin{figure*}
\centering
\includegraphics[width=0.6\textwidth]{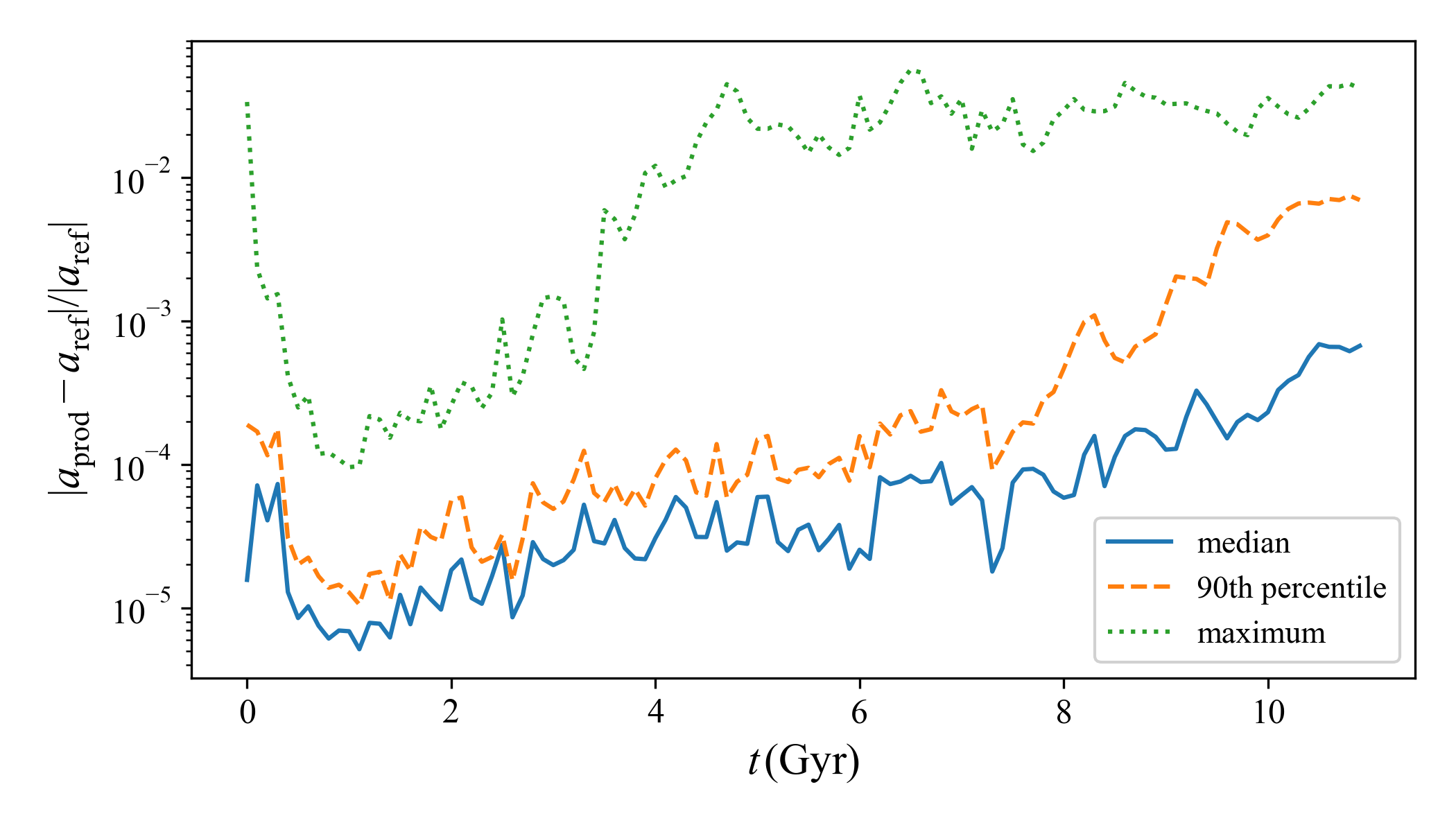}
{\caption{Absolute force accuracy at the evolved particle positions. The production Multipole potential [$N_r=25$, $L_{\max}=20$] is compared with a higher-resolution reference representation [$N_r=100$, $L_{\max}=23$] constructed from the same density snapshots. Curves show the median, 90th-percentile, and maximum relative acceleration differences at representative times.}}
\label{force accuracy}
\end{figure*}
}

\FloatBarrier
\section{Conclusion}

In this study, we used numerical simulations to systematically investigate the heating of stellar clusters in satellite FDM halos, focusing on the regime where the de Broglie wavelength is much larger than the cluster size. Our results have several important implications for constraining FDM using stellar kinematics of ultra-faint dwarf galaxies (UFDs), and they reveal a more nuanced picture than previously appreciated.
Our key findings are the following:
\begin{enumerate}
    \item When the cluster size is much smaller than the de Broglie wavelength, the diffusive heating description is no longer valid. Instead, the cluster undergoes tidal heating, during which its size and velocity dispersion grow exponentially.
    \item The exponential growth terminates when the cluster size becomes comparable to the soliton core size, which is also the scale of the de Broglie wavelength of the halo. Subsequently, the cluster heating enters the diffusive heating stage until the cluster becomes virialized with the halo.
    \item The random walk of the soliton core can enhance the heating.
    \item Tidal stripping of the FDM halo strongly suppresses the heating of the stellar cluster.
\end{enumerate}

Our finding has important implications for constraints
on ultra-light dark matter. The lower bound $m_a\gtrsim3\times10^{-19}\,\mathrm{eV}$ inferred by \cite{dalal2022}  based on a diffusion heating
formula, is valid only when $\lambda_{\rm dB}\ll r_{1/2}$. We show that the time-dependent tidal field heats the stellar component in the opposite regime, with an exponential growth rate approximately given by
\begin{equation}
    k_R\sim 1\left(\frac{m_a}{10^{-22}\mathrm{eV}}\right)\,\mathrm{Gyr}^{-1}.
\end{equation}
Within our adopted idealized halo model, evolving an initially $5$~pc stellar cluster to the observed half-light radii of Segue 1 and Segue 2 gives the illustrative limits $m_a\lesssim2\times10^{-23}\,\mathrm{eV}$ and $m_a\lesssim4\times10^{-23}\,\mathrm{eV}$, respectively. These results reopen the low-mass parameter space that appeared to be excluded by extrapolating the diffusion-heating constraint beyond its regime of validity. These values summarize the outcome of this specific simulation setup and should not be interpreted as robust object-specific constraints. In particular, our halo models were not fitted to the dynamical masses or stellar velocity dispersions of these galaxies. A reliable constraint would require a model that simultaneously reproduces each galaxy's halo mass profile and stellar kinematics and accounts for stellar self-gravity, orbital history, and tidal stripping.

Our principal result is the existence and numerical validation of tidal heating when the stellar system is much smaller than the de Broglie wavelength, together with the transition from exponential to diffusive growth. The illustrative mass limits above should be considered alongside independent observational constraints on FDM (e.g. \cite{Irsic2017,Bar2018,Safar2020,Nadler2020,Laroche2022}) and are particularly sensitive to the halo structure and orbital history.
Reduced soliton mass can relieve the effect of soliton random walk.
Tidal stripping of the satellite halo, as discussed by \cite{Yang2025} can strongly suppress the heating.
When higher-order eigenmodes are removed (simulating tidal stripping), heating is suppressed, to the point of complete suppression when only the static soliton remains. 
The net heating of a UFD therefore depends sensitively on its orbital history, the degree of tidal stripping, and the present-day halo structure. 
This complexity suggests that deriving a universal, one-parameter lower bound on $m_a$ may be overly simplistic. 
Instead, constraints should be interpreted in the context of detailed modeling of individual UFDs, including their orbits and tidal histories, as high eigenmodes may also be excited in a soliton through tidal interactions \cite{widmark2024}.

Our simulations adopt several idealizations that warrant discussion. First, we treat stars as test particles, neglecting self-gravity of the stellar cluster. This is a reasonable approximation for low-mass clusters like UFDs, where the dark matter halo dominates the gravitational potential. However, for more massive clusters, self-gravity could become important and potentially suppress heating \cite{Eber2025}. For compact stellar clusters, self-gravity of stars may dominate at the center and keep system bound \cite{Yang2026}. Second, our halo construction uses a fixed NFW profile with a soliton core; realistic FDM halos may have more complex profiles, especially in the presence of tidal stripping.
\newpage
Third, we have not modeled the orbital motion of the cluster within the host halo; we place the cluster at the halo center, where heating is expected to be strongest. A cluster on an eccentric orbit would experience time-varying tidal fields as it moves through different regions of the halo, potentially leading to more complex heating histories. 

Several extensions of this work are promising. First, full three-dimensional simulations of FDM halos with embedded stellar clusters, including self-gravity and realistic orbits, would provide more accurate heating predictions. Second, observational constraints could be refined by applying our heating models to individual UFDs with well-measured sizes, velocity dispersions, and inferred orbital parameters. Finally, the sensitivity of heating to tidal stripping suggests that the least stripped UFDs — those farthest from the Milky Way or on radial orbits with short pericenter passages — may provide the strongest constraints on FDM.

\begin{acknowledgements}
The authors thank Tomer Yavetz and Oleg Gnedin for helpful discussions.
\end{acknowledgements}

\section*{Data Availability}
The code used in this study is openly available~\cite{Liu2026FDMCode}.

\clearpage
\appendix
\controlsappendix

\clearpage

\bibliography{references}

\end{document}